# Evaluating Search Systems through Field Codes and Classifications: The Case of the First Published Swedish Nursing Study

Christopher Holmberg [1,2]

1. Institute of Health and Care Sciences. University of Gothenburg, Sweden.
2. Department of Psychotic Disorders. Sahlgrenska University Hospital, Sweden.

ABSTRACT
The aim of the study was to assess and compare established search systems and approaches by using the search goal of identifying the first (as in oldest) nursing-related document, with reference to the first Swedish-affiliated document, in Scopus, Web of Science, and PubMed (Medline). In doing so, the objective was to provide concrete examples and illustrate how search systems and field code searches versus classification-based searches differ. The Scopus database, Web of Science database aggregator, and the PubMed (Medline) search engine were used for evaluations. Two different search strategies were compared in each database: one guided by field codes (i.e., text-based), and one guided by pre-existing categorizations within database infrastructures. Findings highlight several factors that are important to consider when formulating and executing a search strategy. The findings illustrate important aspects of search systems and search approaches, namely the publication year, delimitations related to a country or region, and features related to a subject or research area. The findings also highlight the importance of prioritizing between different ideals in retrieval. For example, is it more important to reach the oldest records (publication year), to identify as many records as possible (exhaustiveness), or to make sure they are as subject-relevant as possible (e.g., in terms of authors belonging to the same field)? Secondly, researchers using bibliometric methods to analyze research literature should be more transparent when reporting searches, as different search systems and search approaches yield varied numbers and quality of records. While most bibliometric researchers strive to combine different approaches, this is not always made clear in reporting.

# INTRODUCTION

More and more resources are being invested in research, and the output of research, measured through scientific publications, is on the rise. [1] Thus, bibliometrics, which involves the evaluation and assessment of publication patterns and trends, plays an increasingly important role. [2] Bibliometrics enables researchers and institutions to evaluate the impact of their research output. By analyzing citation patterns and publication trends, one can assess the influence and visibility of specific works, researchers, institutions, or even countries. [3]

Fundamental to bibliometrics and literature reviews are bibliographic search systems that function as digital libraries using various databases. It is important to continuously compare these systems, as their results can differ due to factors such as indexed journals, variations in subject/discipline categorizations, and differences in wording/spelling standardizations for author and affiliation names. [4] There may also be variations in updating speeds of bibliographic records. [5,6] Understanding fundamental academic search concepts is crucial for tailoring search routines and systems appropriately. Successful searches depend on the harmonious alignment of three essential concepts when generating, executing, and retrieving a search: search goals (intention of the search), search systems (e.g., database infrastructure), and heuristics (conceptual operationalization of the search). [7]

This study uses nursing research as a case study to evaluate and compare various databases. Nursing research plays a pivotal role in improving patient outcomes and shaping the future of the nursing profession. Given the extensive influence of nursing research, it is crucial to explore methodological aspects that can affect bibliographic evaluations. Previous research shows variations in search strategies used for bibliometrics in nursing research, differing between word-based searches and searches through pre-defined categorizations. For instance, a bibliometric analysis of nursing research papers in Web of Science affiliated with Taiwan used the words "Taiwan" and "nursing". [3] Another study, aiming to identify gerontological nursing research, used MeSH terms including aged∗, and geriatric nurs∗. In this study, the search field 'Topic' was used, i.e., matching words based on article title, abstract, and keywords. [8] A study aiming to evaluate the coverage of nursing faculty research in Scopus and Web of Science used author names and author affiliations. [9] A large-scale study used the Web of Science Subject area (SU) to denote nursing and evaluated all identified records within a specified year interval. [10]

In this study, search systems and search strategies were evaluated. To refine and contextualize the searches, limitations based on geographic and sociocultural factors were applied, utilizing the research affiliations of document authors within a specific country—Sweden. Besides functioning as a methodological constraint, Sweden has emerged as a significant contributor to nursing research, reflecting a healthcare system that prioritizes person-centered and evidence-based practices. The aim of the study was to assess and compare established search systems and approaches by using the search goal of identifying the first (as in oldest) nursing-related document, with reference to the first Swedish-affiliated document, in Scopus, Web of Science, and PubMed (Medline).

METHODS

This study adopted a comparative bibliometric approach to evaluate search strategies and database infrastructures and assessed these in relation to the objectives of the searches. Bibliometric methods originate from information and library science, generally using a combination of quantitative and qualitative approaches to explore and compare various patterns and trends in published literature. [11] Quantitative elements concern statistically analyzing various metrics (e.g., the number of records, citations) or conducting content or semantic analysis originating in medica research, by using words or sentences (e.g., notable terms used in titles or abstracts). [12] Qualitative approaches might relate to conducting thematic or qualitative content analysis using conceptual definitions and narrative inquiries.

**Search approaches**

Search types or approaches can be considered consisting of three fundamental elements: namely, the goal of the search, the heuristics of searches, and system-related aspects. [7]

The *search goal* is the user's intention in conducting searches, which can be a quick retrieval of information, exploratory searches to clarify concepts and delineations facilitating learning, or a systematic approach that comprehensively retrieves all available literature.

*Heuristics* surrounding the operationalization of searches involve reflecting on choices and their consequences to guide priorities. For example, determining whether it is more important for the most specific record to appear first (i.e., the precision of searches, as in lookup searches) or if retrieval is prioritized (finding the greatest number of closely related records by using cues within the search environment, as in exploratory searches). Alternatively, it may be preferred to strictly use successive fractions (as in systematic searches).

Lastly, *systems* refer to selecting a database based on its functional capabilities and understanding the limitations and biases introduced by algorithms that adjust the order of search results.

**Search systems and databases**

Three system were used for evaluations. Scopus and Web of Science were selected because they are wide-ranging yet curated databases/ aggregators that offers bibliometric metadata of high quality. While it may depend on specific areas or denominations of searches, they generally have a relatively low frequency of database errors and provides coverage of nursing research.[5] The database Medline, through its main search engine PubMed, was also included, as it focuses specifically on health and life sciences including nursing and medicine.

**Nursing: Research and subject area**

In *Scopus*, the term subject areas (SUBJAREA(NURS)) refers to a categorization system that uses All Science Journal Classification (ASJC)). Classification is based on the aims and scope of the source publication and on the content it publishes.

In *Web of Science*, the field tag (WC=Web of Science category) searches the Web of Science categories field within a record. The document category contains the subject category of its source publication. In *PubMed*, the MeSH term "Nursing" (Unique ID: D009729) is assigned by indexing staff at National Library of Medicine based on the full text of a document.

**Sweden: Country**

In *Scopus*, country, or territory (AFFILCOUNTRY) is defined based on the authors' affiliations, using the document's address field.

In *Web of Science*, the field tag (CU=Country/Region) uses the stated countries or regions in the addresses field within a record.

In *PubMed,* affiliations [Affiliation] uses the author address fields to identify a country or region.

All searches underlying the study were conducted on November 24, 2023. For all searches, the author was logged in using his institutional account (University of Gothenburg).

**Data analysis**
In the databases, using various combinations of search queries and denominations, the records were sorted in ascending order, following publication year, so that the oldest records appeared first.

To identify nursing articles, 'nursing' was defined according to the International Council of Nurses' description [13], encompassing independent and collaborative care for individuals across the lifespan, whether they are unwell or in good health, and in various healthcare environments. A nurse, having completed a nursing program, is authorized by the applicable regulatory body to practice nursing within their jurisdiction (e.g., country or state). Professional duties include patient-level activities such as health promotion, illness prevention, and caring for individuals who are unwell, sick, or nearing the end of life. System-level duties encompass ensuring a safe environment for both patients and personnel, conducting research, and education. This definition was considered precise enough to capture the unique features of nursing while being broad enough to accommodate changes in the professional role and healthcare systems over time.

To discern which record was the first published nursing article (in line with the definition) in each search, the bibliographic metadata was screened, that is: article titles, abstracts, keywords, and source journal titles. If any words which used the base "nurs*" appeared (e.g., "nurse", "nurses", "nursing", "nurse education") the full text of the article was evaluated to confirm whether it really was related to nursing.

RESULTS

Results are presented according to the respective search system. First, findings from PubMed (Medline) are presented, followed by Scopus, and finally, Web of Science.

Table 1 shows that a search using the index term "nursing" in All Fields within PubMed yielded nearly 900,000 records. The earliest published document, dating back to 1848, focused on 'nursing' as in caring for infants and small children. The first relevant document (specifically related to nursing) did not emerge until the 19th entry, published in 1860. The gap between the first record overall and the first nursing-related record is 12 publication years and 1-19 places in the search list.

Subsequently, by narrowing the search to documents affiliated with "Sweden," the number of records significantly decreased to 12,092. Surprisingly, the first identified nursing-related document in the Swedish context did not appear until 1987. However, evaluating the search based on authors, 9 out of 10 were researchers in nursing and were based or affiliated with Sweden.

**Table 1.** Searches and queries used in PubMed (Medline) on November 24, 2023.

| Search string | # records | Oldest document (Place in search list, Document title, Publication year) | Oldest nursing-related document (Place in search list, Document title, Publication year) |
|---|---|---|---|
| "nursing"[All Fields] | 890,855 | 1) "Stomatitis Materna." The Sore Mouth of Nursing Women. **1848**.[14] | 19) Florence Nightingale and Her "Notes on Nursing.". **1860**.[15] |
| "nursing"[All Fields] AND Sweden[Affiliation] | 12,092 | 1) Occupational dermatitis in a 10-year material. **1975**.[16] | 3) Staff attitudes before and after the inception of a pediatric oncology unit. **1987**.[17] |
| Evaluation of top 10 most prolific authors: | Nine (9) authors were researchers in nursing (all were Swedish based/affiliated). | | |
| "Nursing"[Mesh] | 264,362 | 1) Obstetrics in Mexico prior to 1600. **1932**.[18] | 8) The nurse; link between patient and doctor. **1945**.[19] |
| "Nursing"[Mesh] AND Sweden | 4,155 | 1) Public health nursing in Sweden. **1946**.[20] | 1) Public health nursing in Sweden. **1946**. |
| Evaluation of top 10 most prolific authors: | All authors were researchers in nursing and Swedish based/affiliated. | | |

While searching in Scopus, initially employing the free text option and truncating 'nurs*' yielded over 3 million records (Table 2). Among them, the earliest nursing-related document was identified; it was published in The Lancet in 1849, addressing the financial aspects of nurses. Even after narrowing the search to only Swedish-affiliated authors, the number of records remained substantial, surpassing 55,000. Upon evaluating the most prolific authors, it was notable that none were researchers in nursing or based in/affiliated with Sweden.

Table 2. Searches and queries used in Scopus on November 24, 2023.

| Search string | # records | Oldest document (Place in search list, Document title, Publication year) | Oldest nursing-related document (Place in search list, Document title, Publication year) |
|---|---|---|---|
| Nurs* | 3,026,952 | 1) Nursery treatment of infants, submitted to Prince Albert. **1842.** [21] | 7. Fees to nurses. **1849**, [22] |
| nurs* AND (LIMIT-TO (AFFILCOUNTRY, "Sweden")) | 55,547 | 1) Variations in the fat content of collected human milk. **1945.** [23] | 3) Nursing in Sweden. **1951**.[24] |
| Evaluation of top 10 most prolific authors: | None (0) of the were researchers in nursing and Swedish based/affiliated. | | |
| SUBJAREA (nurs) | 1,548,374 | 1) Avitaminosis a pathology in dairy calves. **1942.** [25] | 508) Industrial nurse as health educator. **1945.** [26] |
| SUBJAREA (nurs) AND (LIMIT-TO (AFFILCOUNTRY, "Sweden")) | 21,513 | 1) Protein quality. **1955**.[27] | 223) Patients' reactions to impending death. **1979**. [28] |
| Evaluation of top 10 most prolific authors: | Three (3) authors were researchers in nursing and Swedish based/affiliated. | | |

Lastly, using the Web of Science Core Collection, the results show that employing a free text search with the base term 'nurs*' in All Fields yielded almost 850,000 records (Table 3). The first retrieved record was published in 1945, a pattern consistent with nursing-related documents. When narrowing the search to focus on Sweden, the number of records decreased to 21,498. However, the precision, as indicated by the top prolific authors, appeared to be low, as none of the most productive authors were researchers in nursing and were not based or affiliated in Sweden.

A more successful approach seemed to be navigating through the Web of Science categories (WC), selecting 'nursing,' and then filtering to Sweden. In this case, all the most prolific authors were found to be nursing researchers who were based or affiliated in Sweden.

Table 3. Searches and queries used in Web of Science Core collection on November 24, 2023.

| Search string | # records | Oldest document (Place in search list, Document title, Publication year) | Oldest nursing-related document (Place in search list, Document title, Publication year) |
|---|---|---|---|
| nurs* (All Fields) | 843,856 [A] | 1) Basal metabolism of eight nursery school children determined at three month intervals. **1945**. [29] | 2) The benefits of good nursing supervision in institutions for mental defectives. **1945**. [30] |
| nurs* (All Fields) and SWEDEN (Countries/Regions) | 21,498 | 1) Effect of nursery schools on child development. **1970**. [31] | 7) A conceptual framework for nurse staffing management. **1975**. [32] |
| Evaluation of top 10 most prolific authors: | None (0) of the were researchers in nursing and Swedish based/affiliated | | |
| WC=(Nursing) [B] | 321,950 | 1) Home from ETOUSA. **1945**. [33] | 6) Devices to simplify neurological nursing. **1945**. [34] |
| WC=(Nursing) and SWEDEN (Countries/Regions) | 7,941 | 1) The family in the Swedish birth room. **1978**. [35] | 2) Attitudes of nursing staff in general medical wards towards activation of stroke patients. **1982**. [36] |
| Evaluation of top 10 most prolific authors: | All authors were researchers in nursing and Swedish based/affiliated. | | |

[A] This search listed 1,464 records by the author: Nurse, E. University College London Faculty of Mathematical and Physical Sciences. [B] The same results were obtained when using "subject area", SU=(Nursing).

# DISCUSSION

The aim of the study was to assess and compare established search systems and approaches by using the search goal of identifying the first (as in oldest) nursing-related document, with reference to the first Swedish-affiliated document, in Scopus, Web of Science, and PubMed (Medline).

Compiling the results from the different search systems, to enable more straightforward comparisons across systems, Table 4 was generated.

Table 4. Comparing search systems in terms of searches identifying Swedish nursing research.

| Search system: Field Code or Classification | # Records | First relevant record (publication year) | Relevant authors (out of 10 most prolific) |
|---|---|---|---|
| PubMed: Field code | 12,092 | Staff attitudes before and after the inception of a pediatric oncology unit. **1987**. [17] | 9/10 |
| PubMed: Classification | 4,155 | Public health nursing in Sweden. **1946**.[20] | 10/10 |
| Scopus: Field code | 55,547 | Nursing in Sweden. **1951**.[24] | 0/10 |
| Scopus: Classification | 21,513 | Patients' reactions to impending death. **1979**. [28] | 3/10 |
| Web of Science: Field code | 21,498 | A conceptual framework for nurse staffing management. **1975**. [32] | 0/10 |
| Web of Science: Classification | 7,941 | Attitudes of nursing staff in general medical wards towards activation of stroke patients. **1982**. [36] | 10/10 |

**Differences in search systems**

In *PubMed*, searches using field codes or classifications resulted in 9/10 and 10/10 relevant authors, respectively, indicating that searches were within the topic regardless of the approach used. However, the first relevant article using field codes was published in 1987, while the first relevant article identified using classifications was published in 1946. This suggests that not the same articles were found. As such, it indicates that the level of precision (accuracy in terms of aligning with search goal definition) was high.

In *Scopus*, searches using field codes or classifications resulted in 0/10 and 3/10 relevant authors, respectively, indicating that searches were not truly within the topic regardless of the approach used. In *Web of Science*, searches using field codes or classifications resulted in 0/10 and 10/10 relevant authors, respectively, suggesting that searches using classifications were far more accurate and precise in terms of finding subject-specific authors.

Based on the results, searching PubMed and Web of Science using classifications would provide an ideal combination, as both searches resulted in relevant authors (high level of precision), a different number of total records (recall capacity), as well as unique documents which constituted the oldest/first (which indicate system variations in samples). These results, showing clear differences between search systems, align with other evaluations of search systems that has used other indicators and approaches for comparisons. [6]

**Limitations**
This evaluation is constrained by certain limitations. Firstly, the evaluation relied on only three search systems. While these are regularly used and often recommended, considering the emergence of the number of databases, it's noteworthy that some studies have evaluated up to n=56. [5] Thus, the limited number of included databases in this study should be taken into account. Secondly, there are specific factors associated with the search systems themselves that complicate comparisons between systems. Most search systems not only update their databases but also enhance their search functionalities. Consequently, the results from evaluations in this study might evolve over time. Additionally, as mentioned earlier, there could be variations between databases in terms of their updating speeds of bibliographic records. [5,6] This variability could impact the search results, introducing a time-dependent element.

In summary, this study can only demonstrate that different search approaches and systems yield varying results in terms of precision and retrieval under the specified test criteria. It cannot be ruled out that a strategy deemed successful or unsuccessful in this study might yield different outcomes in other evaluations under different circumstances.

**Conclusions and implications**
A pivotal factor that can significantly hinder the quality of our work is the misconception that our current academic search practices are of sufficient standard. This mindset assumes that the systems that are routinely employed, and the current information retrieval habits are sufficient for effective and efficient information retrieval. However, the process of searching, a central element in research, demands acquired skills, thoughtful consideration, and planning. It is imperative to recognize that the 'where' and 'how' of searches profoundly influence what is discovered and overlooked, our conclusions, and the evidence-based recommendations that are proposed. Enhancing academic search proficiency, in nursing and otherwise, not only elevates the quality of scientific endeavors but also contributes to combating information epidemics. Therefore, there is much to be gained by refining the day-to-day academic searching practices of millions of researchers worldwide.

This study contributes to information retrieval practices by aligning with previous large-scale studies, [6] while adopting a narrower and field-oriented goal for searches. The intention is to raise awareness of the importance of making specific strengths and limitations of search systems and approaches transparent and reproducible. In this study, the search systems PubMed (Medline) and Web of Science, when using classifications, provided the most accurate results. Combining these two systems would be ideal for the goal of the current study – were it to be conducted in full. If one could only select one however, since PubMed allows access to all datasets returned from a single search, and Web of Science's Core Collection might vary substantially in scope (including different indices based on institutional subscriptions), [6] PubMed appears to be the more favorable system in this context.